\documentclass[twocolumn,epjc3]{svjour3}          

\usepackage{amsmath} 

\usepackage{amssymb} 
\usepackage{graphicx}

\newcommand\be{\begin{equation}}
\newcommand\ee{\end{equation}}
\newcommand\ba{\begin{eqnarray}}
\newcommand\ea{\end{eqnarray}}

\newcommand\e{{\rm e}}
\newcommand\de{\delta}

\newcommand\al{\alpha}
\newcommand\bt{\beta}
\newcommand\pa{\partial}
\newcommand\Om{\Omega}
  \newcommand{\h}{{\cal H}}

\renewcommand{\(}{\left(}
\renewcommand{\)}{\right)}
\renewcommand{\[}{\left[}
\renewcommand{\]}{\right]}

\newcommand{\lm}{\lambda}
\newcommand{\sig}{\sigma}
\newcommand{\om}{\omega}

\newcommand{\m}{{\cal M}}
\newcommand{\dm}{{c}}
\newcommand{\dmp}{{m {\rm p}}}

\newcommand\Cp{C_{,\phi}}
\newcommand\Dp{D_{,\phi}}

\newcommand\DX{D_{,X}}

\newcommand\ja{{\rm J}}

\RequirePackage[T1]{fontenc}

\smartqed  

\RequirePackage{graphicx}

\RequirePackage{mathptmx}      
\RequirePackage{flushend}
\RequirePackage[numbers,sort&compress]{natbib}
\RequirePackage[colorlinks,citecolor=blue,urlcolor=blue,linkcolor=blue]{hyperref}

\journalname{Eur. Phys. J. C}

\begin{document}

\title{Dynamics of the universe with disformal coupling between the dark sectors}

\author{Khamphee Karwan\thanksref{addr1,e1}
        \and
        Stharporn Sapa\thanksref{addr1,e2}}

\thankstext{e1}{e-mail: khampheek@nu.ac.th}
\thankstext{e2}{e-mail: stharporns57@email.nu.ac.th}

\institute{The Institute for Fundamental Study ``The Tah Poe Academia Institute'', Naresuan University,
Phitsanulok, 65000 Thailand\label{addr1}}

\date{Received: date / Accepted: date}

\maketitle

\begin{abstract}
We use the dynamical analysis to study the evolution of the universe at late time for the model in which the interaction between dark energy and dark matter is inspired by disformal transformation. We extend the analysis in the existing literature by supposing that the disformal coefficient depends both on the scalar field and its kinetic terms. We find that the dependence of the  disformal coefficient on the kinetic term of scalar field leads to two classes of the scaling fixed points that can describe the acceleration of the universe at late time. The first class exists only for the case where the disformal coefficient depends on the kinetic terms. The fixed points in this class are saddle points unless the slope of the conformal coefficient is sufficiently large. The second class can be viewed as the generalization of the fixed points studied in the literature. According to the stability analysis of these fixed points, we find that the stable fixed point can take two different physically relevant values for the same value of the parameters of the model. These different values of the fixed points can be reached for different initial conditions for the equation of state parameter of dark energy. We also discuss the situations in which  this feature disappears.
\end{abstract}

\section{Introduction}

The observed cosmic acceleration at late time is the one of the most important mystery in the universe \cite{SNIa,Planck}.
This  phenomena may be explained by introducing unknown form of energy to govern the dynamics of the late-time universe \cite{sahni,Copeland:06}.
For the simplest case, this unknown form of energy, dubbed dark energy, is supposed to be in the form of evolving scalar fields.
In general, viable dark energy models should have a mechanism to alleviate the coincidence problem,
which is the problem why the energy density of dark energy and matter are comparable in magnitude at present although they evolve independently throughout the whole evolution of the universe \cite{coin1,coin2,coin3}.
A possible assumption for alleviating the coincidence problem is based on the introduction of the interaction between dark energy and dark matter.
Various phenomenological form  of the interaction between dark energy and dark matter have been proposed and investigated in literature \cite{kk:08,inter1,inter2,inter3}.
Interestingly, it has recently been shown that models of dark energy in which the interaction between dark energy and dark matter is assumed can satisfy the bound on the Hubble parameter at redshift  2.34 from BOSS data, while the $\Lambda$CDM model predicts too large Hubble parameter at this redshift \cite{binw:14}.

In addition to the phenomenological models of the interaction between dark energy and dark matter, the models of the interaction between dark energy and dark matter can be constructed from the frame transformation of the theory of gravity. Applying the conformal transformation to some classes of scalar-tensor theories, one obtains the coupling terms between dark energy and dark matter in the Einstein frame in which the gravity action takes the Einstein-Hilbert form \cite{conform,cft, Clifton:11}. The cosmological consequences of the the interaction between dark energy and dark matter due to the conformal transformation have been investigated in \cite{conformc,confor inter}. However, in order to transform general scalar-tensor theories to Einstein frame, we need transformations that are more general than the conformal transformation. It has been shown that subclasses of the GLPV theory which is the generalization of the Horndeski theory can be transformed to the Einstein frame using the disformal transformation defined as \cite{Bekenstein:92, glpv, van de Bruck}
\begin{equation}
\label{gbar-gen}
\bar{g}_{\mu\nu} = C(\phi)g_{\mu\nu} + D(\phi,X)\phi_{,\mu}\phi_{,\nu}\,,
\end{equation}
where subscript ${}_,$ denotes partial derivatives, and \\
$X = - \frac{1}{2}\pa_\mu\phi\pa^\mu\phi $ is the kinetic energy of scalar field. Here, the conformal coefficient $C$ depend only on scalar field $\phi$, while the disformal coefficient $D$ can depend both on $\phi$ and its kinetic term $X$. In the case where $D$ depends only on the scalar field, the above transformation provides relations among some pieces of Lagrangian in the Horndeski theory, but cannot generate a piece of the GLPV Lagrangian from the Horndeski theory \cite{Bettoni:13}. However, it has been shown that if $C$ also depends on $X$, the application of the transformation to GLPV action can generate terms that do not belong to the GLPV theory \cite{glpv} and therefore these terms might be a cause of Ostrogradski's instibilities in the theories. Nevertheless, according to the discussion in \cite{Zumalacarregui:13}, the Ostrogradski's instabilities can be eliminated by hidden constraint in some cases.

The cosmological consequences of the interaction between dark energy and dark matter due to the disformal transformation, called disformal coupling between dark energy and dark matter, has been studied in various aspects for the case where the conformal and disformal coefficients depend on the field $\phi$ only. It has been shown in \cite{Sakstein:141} that the disformal coupling between dark energy and dark matter leads to a new stable fixed point compared with the case of conformal coupling, and the cosmological parameters at this fixed point can satisfy the observational bounds. The metric singularity of the new fixed point found in \cite{Sakstein:141} presents the phantom behaviour in the Jordan frame \cite{Sakstein:15}. The influences of the disformal coupling on the observational quantities such as the CMB and matter power spectra have been investigated in \cite{David:12, David:121, vandeBruck:15}. In this work, we study the disformal coupling between dark energy and dark matter for more general case where the disformal coefficients depends on both $\phi$ and its kinetic terms $X$. The physical motivation for such disformal coupling is related to the frame transformation among the general scalar-tensor theories presented above. Our aims are to study how the kinetic-dependent disformal coupling influences the evolution of the universe at late time by finding and analyzing the physically relevant fixed points of the model, rather than search for all possible fixed points of the model. We will show in the following sections that there are features arising only for the case where the disformal coefficient depends on both $\phi$ and $X$.
 
In section \ref{sec:2}, the evolution equations for dark energy and dark matter with disformal coupling are presented in the covariant form.
The autonomous equations for this model of dark energy are computed in section \ref{sec:3}, and the evolution of the late-time universe is studies using the dynamical analysis in section \ref{sec:4}. The conclusions are given in section \ref{sec:5}.

\section{Disformal coupling between dark energy and dark matter}
\label{sec:2}

In this section, we derive the disformal coupling between scalar field and matter arising from the general disformal transformation defined in eq.~(\ref{gbar-gen}). From the metric in eq.~(\ref{gbar-gen}), we have
\ba
\bar{g}^{\mu\nu} 
= \frac{1}{C}g^{\mu\nu} 
- \frac{D \phi^{,\mu}\phi^{,\nu}}{C^2 - 2CDX}
\,,
\ea
In order to study coupling between dark energy and dark matter due to disformal transformation, we suppose that the field $\phi$ in the disformal transformation plays a role of dark energy, and therefore the interaction between dark energy and dark matter can occur when the Lagrangian of dark matter depends on metric $\bar{g}_{\mu\nu}$ defined in eq.~(\ref{gbar-gen}). Thus we write the action for gravity in terms of metric $g_{\mu\nu}$ and write the action for the dark matter in terms of $\bar{g}_{\mu\nu}$ as
\begin{align}
S & =\int d^4x \{ \sqrt{-g} \big[\frac{1}{2}R + P(\phi,X)+ \mathcal{L}_\m(g_{\mu\nu})\big] \nonumber \\
& + \sqrt{ -\bar{g}}\mathcal{L}_\dm(\bar{g}_{\mu\nu}, \psi, \psi_{,\mu}) \}
\label{action}
\end{align}
where we have set $1/\sqrt{8\pi G} = 1$, $P(\phi,X) \equiv X - V(\phi)$ is the Lagrangian of the scalar field, $V(\phi)$ is the potential of the scalar field, $\mathcal{L}_\dm$ is the Lagrangian of dark matter and $\mathcal{L}_\m$ is the Lagrangian of ordinary matter including baryon and radiation. Varying this action with respect to $g_{\al\bt}$, we get
\be
G^{\al\bt} = T^{\al\bt}_\phi + T^{\al\bt}_\dm + T^{\al\bt}_\m\,,
\label{eom-g1}
\ee
where $G^{\al\bt}$ is the Einstein tensor computed from $g_{\mu\nu}$,
and the energy momentum tensor for scalar field and matter are defined in unbarred frame as
\begin{align}
T^{\mu\nu}_{\phi} &\equiv \frac{2}{\sqrt{-g}}\frac{\delta(\sqrt{-g}P(\phi,X))}{\delta g_{\mu\nu}}, T^{\mu\nu}_{\m} \equiv \frac{2}{\sqrt{-g}} \frac{\delta(\sqrt{-g} \mathcal{L}_\m)}{\delta g_{\mu\nu}}
\label{tmn-p}\\
T^{\mu\nu}_{\dm} &\equiv \frac{2}{\sqrt{-g}}\frac{\delta\(\sqrt{-\bar{g}}\mathcal{L}_{\dm}\)}{\delta g_{\mu\nu}}\,.
\label{tmn-m}
\end{align}
Using these definitions of the energy momentum tensor and the conservation of the energy momentum tensor of ordinary matter, we have $\nabla_\al (T^{\al\bt}_\phi + T^{\al\bt}_{\dm}) =0$ due to $\nabla_\al G^{\al\bt}=0$. However, we see that the energy momentum tensors of dark energy and dark matter do not separately conserve because the Lagrangian of dark matter depends on field $\phi$. On the other hand, since $\bar{g}_{\al\bt}$ does not depend on $\psi$, the energy momentum tensor of dark matter is conserved in the barred frame such as
\be
\bar\nabla_\al \bar{T}^{\al\bt}_\dm =0\,,
\label{conserve-tb}
\ee
where $\bar\nabla_\al$ is defined from barred metric, and the energy momentum tensor in the barred frame is related to that in the unbarred frame defined in eq.~(\ref{tmn-m}) as
\begin{align}
T^{\al\bt}_\dm &= \frac{\sqrt{-\bar{g}}}{\sqrt{-g}}\, \frac{\de \bar{g}_{\rho\sig}}{\de g_{\al\bt}}\, \frac{2}{\sqrt{-\bar{g}}} \frac{\delta\(\sqrt{-\bar{g}}\mathcal{L}_{\dm}\)}{\delta \bar{g}_{\rho\sig}} \nonumber\\
& = \frac{\sqrt{-\bar{g}}}{\sqrt{-g}}\frac{\de \bar{g}_{\rho\sig}}{\de g_{\al\bt}}\bar{T}^{\rho\sig}_\dm\,.
\label{bar-unbar}
\end{align}
To compute the interaction terms between scalar field and dark matter, we vary the action with respect to $\phi$ as
\be
\de S = 
\underbrace{\int dx^4 \sqrt{-g} \frac{\de P}{\de\phi}\de\phi}_{S_\phi} + \underbrace{\int d^4x \frac{\de(\sqrt{-\bar{g}} \mathcal{L}_{\dm})}{\de\phi}\de\phi}_{S_\dm} = 0\,.
\label{desdephi}
\ee
One can show that
\be
S_\phi
=
\int d^4x \sqrt{-g}\(
\phi^{,\mu}_{;\mu} - V_{,\phi}
\)\de\phi,
\label{dsphi}
\ee
where ${}_;$ denotes the covariant derivative and $V_{,\phi} = \pa V / \pa \phi$,
and
\ba
S_\dm &=& \int d^4x \frac{\de(\sqrt{-\bar{g}} \mathcal{L}_{\dm})}{\de \bar{g}_{\al\bt}}\frac{\de\bar{g}_{\al\bt}}{\de \phi}\de\phi \nonumber \\
&=&\int d^4x \Big\{\frac{\sqrt{-\bar{g}}}2 \[ \bar{T}^{\al\bt}_\dm\(\Cp g_{\al\bt} + \Dp\phi_{,\al}\phi_{,\bt}\)\] \nonumber \\
&& -\sqrt{-g}\nabla_\bt (\frac{\sqrt{-\bar{g}}}{\sqrt{-g}} \bar{T}^{\al\bt}_\dm D \phi_{,\al}) \nonumber \\
&& + \frac{\sqrt{-g}}2 \nabla_\om \Big[\phi^{,\om} \frac{\sqrt{-\bar{g}}}{\sqrt{-g}} \bar{T}^{\al\bt}_\dm\(\DX \phi_{,\al}\phi_{,\bt}\)\Big]\Big\}\,\de\phi
\label{dsm}
\ea
Combining eq.~(\ref{dsphi}) with the above equation, we obtain the evolution equation for scalar field $\phi$,
\ba
\phi^{,\al}_{;\al} - V_{,\phi} 
&=& \nabla_\bt\(\frac{\sqrt{-\bar{g}}}{\sqrt{-g}} \bar{T}^{\al\bt}_\dm D \phi_{,\al}\) \nonumber \\
&&- \frac 12 \frac{\sqrt{-\bar{g}}}{\sqrt{-g}} \bar{T}^{\al\bt}_\dm\(\Cp g_{\al\bt} + \Dp\phi_{,\al}\phi_{,\bt}\) \nonumber\\
&& - \frac 12 \nabla_\om\[\phi^{,\om} \frac{\sqrt{-\bar{g}}}{\sqrt{-g}}\bar{T}^{\al\bt}_\dm\DX \phi_{,\al}\phi_{,\bt}\]
\equiv Q\,.
\label{gkg-q1}
\ea
Multiplying the above equation by $\phi^{,\lm}$, we get
\be
Q\phi^{,\lm} 
=
\nabla_\al T^{\al\lm}_\phi = - \nabla_\al T^{\al\lm}_\dm\,.
\label{dt-q}
\ee
In order to write the barred quantities in the interaction term $Q$ in terms of unbarred quantities, we write eq.~(\ref{bar-unbar}) as
\be
T^{\al\bt}_\dm 
= \(C \de^\al_\rho\de^\bt_\sig - \frac 12 \DX \phi^{,\al}\phi^{,\bt}\phi_{,\rho}\phi_{,\sig}\)
\ja\bar{T}^{\rho\sig}_\dm\,,
\label{bar-unbar-cd}
\ee
where $\ja \equiv \sqrt{-\bar{g}} / \sqrt{-g}$.
Hence, we get
\ba
T_\dm 
&=& g_{\al\bt} T^{\al\bt}_\dm 
= \ja C g_{\rho\sig} \bar{T}^{\rho\sig}_\dm
+ \ja\DX X \phi_{,\rho}\phi_{,\sig}\bar{T}^{\rho\sig}_\dm\,,
\\
T_\dmp &\equiv&
\phi_{,\al}\phi_{,\bt} T^{\al\bt}_\dm 
= 
\ja\(C - 2 \DX X^2\)\phi_{,\rho}\phi_{,\sig} \bar{T}^{\rho\sig}_\dm\,,
\ea
and therefore
\ba
g_{\al\bt}\bar{T}^{\al\bt}_\dm
&=&
\frac{C T_\dm - D_{, X} X \bigl(T_\dmp + 2 T_\dm X\bigr)}{C J \bigl(C - 2 D_{, X} X^2\bigr)}\,,
\label{gbt}\\
\phi_{,\al}\phi_{,\bt}\bar{T}^{\al\bt}_\dm
&=&
\frac{T_\dmp}{J \bigl(C - 2 D_{, X} X^2\bigr)}\,.
\label{ppbt}
\ea
These relations yield
\ba
\bar{T}^{\al\bt}_\dm=\frac{T^{\al\bt}_\dm}{C \ja} + \frac{D_{, X} \phi{}^{,\al} \phi{}^{,\bt}}{2 C \ja (C - 2 D_{, X} X^2)}T_\dmp\,.
\label{t2bt}
\ea
Inserting eqs.~(\ref{gbt}), (\ref{ppbt}) and (\ref{t2bt}) into eq.~(\ref{gkg-q1}) we can write $Q$ as
\ba
FQ &=& C [-2 D F_1 F_{\text{1,} \phi} + C F_1 (- D_{, \phi} + F_{\text{2,} X \phi}) + D_{, X} (C_{, \phi} F_1 \nonumber\\
&& - 2 F_{\text{1,} \phi} F_2) X ] T_\dmp - C C_{, \phi} F_1 (C - 2 D_{, X} X^2) T_\dm \nonumber\\
&& - C D_{, X} F_1 F_2 T_\dmp \Box\phi + 2 C D F_1^2 \Theta_1 + 2 C D_{, X} F_1^2 \Theta_2 \nonumber\\
&& + 2 C D F_1^2 \Theta_3 - C D_{, X} F_1 F_2 \Theta_4 - C (D_{,\text{XX}} F_1 F_2 \nonumber \\
&& - D_{, X} F_{\text{1,X}} F_2 + D_{, X} F_1 F_{\text{2,X}}) T_\dmp \Theta_5\,,
\label{q-ub}
\ea
where $F_1 \equiv C - 2 D_{,X} X^2$,\quad $F_2 \equiv C + 2 D X$,\quad $F \equiv 2 C^2 F_1^2$ and
\ba
\Theta_1 &=& 
\phi_{;\al\bt} T^{\al\bt}_\dm\,,
\hspace{0.2cm}\Theta_2 =
\phi_{,\al} X_{,\bt} T^{\al\bt}_\dm\,,
\hspace{0.2cm}\Theta_3 =
\phi_{,\al} \nabla_\bt T^{\al\bt}_\dm\,,
\label{th1}\\
\Theta_4 &=&
\phi^{,\al} \nabla_\al T_\dmp\,,
\hspace{0.2cm}\Theta_5 =
\phi^{,\al} X_{,\al}\,.
\label{th5}
\ea
The form of this interaction terms can reduce to that in \cite{David:12, vandeBruck:15} when $\lm_3 = 0$.

\section{Dynamical equations}
\label{sec:3}

\subsection{Evolution equations for the FLRW universe}

We now compute the evolution equations for all matter components in the spatially flat FLRW universe. Using the perfect fluid model for radiation and matter as well as the FLRW line element in the form
\be
ds^2 = a^2(\tau) \(- d\tau^2 + \de_{ij}dx^i dx^j\)\,,
\label{ds-flrw}
\ee
the component (00) of eq.~(\ref{eom-g1}) yields
\be
\h^2 \equiv \(\frac{a'}{a}\)^2 = \frac 13 \[a^2\(\rho_r + \rho_b + \rho_c\)+ \frac 12 (\phi')^2 + a^2V(\phi)\]\,,
\label{h2}
\ee
where a prime denotes derivative with respect to conformal time $\tau$,
$\rho_r$, $\rho_b$ and $\rho_c$ are the energy density of radiation, baryon and dark matter respectively. Furthermore, the interaction terms $Q$ in eq.~(\ref{q-ub}) becomes
\ba
FQ &=&
2 C \phi' \rho_c' F_1 \bigl(D F_1 + D_{, X} F_2 X\bigr) + a^2 C \rho_c [-4 F_{\text{1,} \phi} X \bigl(D F_1 \nonumber\\  
&& + D_{, X} F_2 X\bigr) + C F_1 \bigl(C_{, \phi} + 2 (- D_{, \phi} + F_{\text{2,} X \phi}) X\bigr)] \nonumber \\
&& + 4 C \phi' \h \rho_c [D F_1^2 - X \bigl(D_{, \text{XX}} F_1 F_2 X\nonumber \\
&& + D_{, X} (F_1^2 - F_{\text{1,X}} F_2 X + F_1 F_{\text{2,X}} X)\bigr)] \nonumber \\
&&+ 2 C \phi'' \rho_c [D F_1^2 + X \bigl(2 D_{, \text{XX}} F_1 F_2 X \nonumber \\
&&+ D_{, X} (2 F_1^2 + 3 F_1 F_2 - 2 F_{\text{1,X}} F_2 X + 2 F_1 F_{\text{2,X}} X)\bigr)]\,. 
\label{q0-start}
\ea
From now on, we will use $X \equiv (\phi')^2 / (2 a^2)$. Inserting this expression for the interaction terms into eq.~(\ref{dt-q}), we can write $\rho_c'$ as
\ba
F_c\rho_c' 
&=&
a^2 C \phi' \rho_c \Big[-4 F_{\text{1,} \phi} X (D F_1 + D_{, X} F_2 X) + C F_1 \big(C_{, \phi} \nonumber \\
&& + 2 ( - D_{, \phi} + F_{\text{2,} X \phi}) X \big)\Big] + 2 C \phi '' \phi ' \rho_c \Big[D F_1^2 \nonumber \\
&& + X \big(2 D_{, \text{XX}} F_1 F_2 X + D_{, X} (2 F_1^2 + 3 F_1 F_2 - 2 F_{\text{1,X}} F_2 X  \nonumber \\
&& + 2 F_1 F_{\text{2,X}} X)\big)\Big] - 2 C \h \rho_c \Big[3 a^2 C F_1^2+ 4 a^2 X \Big(- D F_1^2 \nonumber\\ 
&& + X \big(D_{, \text{XX}} F_1 F_2 X + D_{, X} (F_1^2 - F_{\text{1,X}} F_2 X + F_1 F_{\text{2,X}} X)\big)\Big)\Big] \nonumber \\
\label{rhomp1}
\ea
where $F_c \equiv 2 a^2 C F_1 [C F_1 - 2 X (D F_1 + D_{, X} F_2 X)]$.
Substituting this expression for $\rho_c'$ into eq.~(\ref{q0-start})
and using eq.~(\ref{gkg-q1}), we get
\ba
\phi'' && + 2 \h \phi' + V_{,\phi}a^2 = F_\phi^{-1} \rho_c \Bigl\{ 6 \phi' \h \bigl(F_1^2 F_d + F_1 F_2 F_{\text{d,X}} X \nonumber \\
&& - 2 D_{, X} F_{\text{d3}} X^2\bigr) + a^2 C \rho_c \bigl(- C_{, \phi} F_1 + 2 (D_{, \phi} F_1 -  F_1 F_{\text{2,} X \phi} \nonumber \\
&& + 2 F_{\text{1,} \phi} F_{\text{d1}}) X\bigr) + 2 a^2 V_{, \phi} \bigl(F_1^2 F_d + F_1 F_2 F_{\text{d,X}} X \nonumber \\
&& - 2 D_{, X} F_{\text{d3}} X^2\bigr)\Bigr\}\equiv - Q_0\,,
\label{defq0}
\ea
where 
\ba
F_\phi &=& 2 \big[C F_1 (F_1 - 2 F_{\text{d1}} X) + \rho_c (F_1^2 F_d + F_1 F_2 F_{\text{d,X}} X \nonumber \\
&& - 2 D_{, X} F_{\text{d3}} X^2)\big]
\label{fphi}
\ea
and
\ba
F_d &\equiv& D + 2 X D_{,X}\,,
\qquad\quad\quad
F_{d1} \equiv D + D_{,X} X\,,
\nonumber\\
F_{d2} &\equiv& D F_{1,X} + D_{,X}X F_{2,X}\,,
\quad
F_{d3} \equiv F_{1,X}F_2 - F_{2,X} F_1\,.
\ea
The evolution equations for $\rho_r$ and $\rho_b$ can be computed from the conservation of their energy yielding
\ba
\rho_r' = - 4\h \rho_r\,,
\qquad
\rho_b' = - 3\h \rho_b\,.
\label{rrprbp}
\ea

\subsection{Autonomous equations}

In order to study the evolution of the universe  for the disformal coupled model of dark energy, we analyze solutions for the evolution equations presented in the previous section using dynamical analysis. For concreteness, we derive the autonomous equations using the conformal coupling, disformal coupling and the scalar field potential of the form 
\be
C = C_0 \e^{\lm_1 \phi}\,,
\quad
D = M^{-4 - 4 \lm_3} \e^{\lm_2 \phi} X^{\lm_3}\,,
\quad
V  = M_v^4 \e^{\lm_4\phi}\,,
\label{def-cd}
\ee
where $\lm_1$, $\lm_2$, $\lm_3$, $\lm_4$ and $C_0$ are the dimensionless constant parameters, while $M$ and $M_v$ are the constant parameters with dimension of mass. Here, we extend the analysis in the literature by supposing that the disformal coefficient $D$ also depends on the kinetic term $X$ through $X^{\lm_3}$ which is the simplest extension. Using the following dimensionless variables,
\ba
\Omega_c &\equiv& \frac{a^2\rho_c}{3 \h^{2}}\,,
\quad
\Omega_r \equiv \frac{a^2\rho_r}{3 \h^{2}}\,,
\quad
\Omega_b \equiv \frac{a^2\rho_b}{3 \h^{2}}\,,
\nonumber\\
x_1^2 &\equiv & \frac{\phi'^2}{6 \h^{2}}\,,
\quad 
x_2 \equiv \frac{a^2 V}{3 \h^{2}}\,,
\quad
x_3 \equiv \frac{D \h^{2}}{a^2 C}\,,
\label{def-dyna}
\ea
we can write eqs.~(\ref{defq0}) and (\ref{rrprbp}) in the form of autonomous equations as
\ba
\frac{d x_1}{d N} &=& \frac {1}{q} \{ x_1 (\Om_r + 3 x_1^2 - 3 x_2 +1) \big[18 \lambda_3 x_1^2 x_3^2 \big((3 \lambda_3+1) x_1^2 \nonumber\\
&&
- (\lambda_3+1) (\Om_b + \Om_r + x_2 - 1)\big) + 3 x_3 \big(- 2 \lambda_3^2 (\Om_b + \Om_r \nonumber\\ 
&&
+ x_1^2 + x_2 - 1) - (3 \lambda_3+1) (\Om_b + \Om_r + 3 x_1^2 + x_2 - 1)\big)\nonumber\\
&&
+ 1\big] - 2 \big(\sqrt{3/2} \lambda_4 x_2 + 2 x_1\big) \big[18 \lambda_3 x_1^2 x_3^2 \big((3 \lambda _3 + 1) x_1^2  \nonumber\\ 
&&
- (\lambda_3 + 1) (\Om_b + \Om_r + x_2 - 1)\big) + 3 x_3 \big(- 2 \lambda_3^2 (\Om_b + \Om_r   \nonumber\\
&&
+ x_1^2 + x_2 - 1) - (3 \lambda_3 + 1) (\Om_b + \Om_r + 3 x_1^2 + x_2 - 1)\big) \nonumber\\ 
&&
+ 1\big] -\sqrt{3/2} \big[6 x_3 \big(\lambda_3 (6 x_3 x_1^2 + 2) + 1\big) \big(- \lambda_2 x_1^2 \nonumber\\ 
&&
+ \sqrt{6} (\lambda_3 + 1) x_1 + (\lambda_3 +  1) \lambda_4 x_2 \big)+ \lambda_1 \big(6 (3 \lambda_3 + 2) x_1^2 x_3 \nonumber\\
&&
- 1\big)\big](\Om_b + \Om_r + x_1^2 + x_2 - 1)\},
\label{x1p}\\
\frac{d x_2}{d N}
&=&
x_2 \big(\Om_r + \sqrt{6} \lambda_4 x_1 + 3 x_1^2-3 x_2 + 3\big)\,,
\label{x2p}\\ 
\frac{d x_3}{d N}
&=& 
- x_3 [3 \lambda_3 + (\lambda_3+ 1) \Om_r - 2 \frac{\lambda_3}{x_1}\frac{d x_1}{d N} + 3 (\lambda_3 + 1) x_1^2 \nonumber\\
&&
+ \sqrt{6} (\lambda_1 - \lambda_2) x_1-3 (\lambda_3 + 1) x_2+3]\,,
\label{x3p}\\
\frac{d \Om_b}{d N} 
&=&
\Om_b (\Om_r + 3 x_1^2 -3 x_2)\,,
\label{omp}\\
\frac{d \Om_r}{d N} 
&=&
\Om_r \left(\Om_r + 3 x_1^2-3 x_2-1\right)\,,
\ea
where $N \equiv \ln a$ and 
\ba
q & \equiv &
2 [18 \lambda_3 x_1^2 x_3^2 ((3 \lambda_3 + 1) x_1^2 - (\lambda_3 + 1) (\Om_b + \Om_r + x_2 - 1)) \nonumber\\
&&
+ 3 x_3 (- 2 \lambda_3^2 (\Om_b + \Om_r +x_1^2+x_2-1)-(3 \lambda_3 + 1) (\Om_b + \Om_r \nonumber\\
&&
+ 3 x_1^2 + x_2 - 1)) + 1]
\ea
The evolution of the universe is completely described by these autonomous equations and the constraint equation which is obtained from eq.~(\ref{h2}) as
\be
\label{constraint}
1= x_1^2 + x_2 + \Om_c + \Om_b + \Om_r\,.
\ee
In order to derive the above autonomous equations, we also use
\be
\frac 1{\h^2} \frac{d \h}{d \tau} = 
\frac 12 \(3 x_2 - 3 x_1^2 - 1 - \Om_r\)\,,
\label{hprime}
\ee
which can be obtained by differentiating the constraint equation with respect to $N$. From the above equation, we see that
\be
\frac{d^2 a}{dt^2}
=
\frac{\h^2}{2 a} \(3 x_2 - 3 x_1^2 - 1 - \Om_r\)\,,
\label{addot}
\ee
where $t$ is the cosmic time.

\newcommand{\xc}{x_{1f}}
\newcommand{\yc}{x_{2f}}
\newcommand{\zc}{x_{3f}}
\newcommand{\odc}{\Om_{df}}
\newcommand{\wdc}{w_{df}}
\newcommand{\gac}{\gamma_{f}}
\newcommand{\ode}{\Om_{d}}
\newcommand{\wde}{w_{d}}
\newcommand{\ip}{{\rm I}^+}
\newcommand{\im}{{\rm I}^-}
\newcommand{\iip}{{\rm II}^+}
\newcommand{\iim}{{\rm II}^-}

\section{Dynamical analysis}
\label{sec:4}

Here, we concentrate on dynamics of the universe at late time, so that we ignore the contribution from radiation density in the autonomous equations. Moreover, we are mainly interested in the physical fixed points that correspond to the acceleration of the universe at late time.

The fixed points of the autonomous equations can be obtained by setting the LHS of eqs.~(\ref{x1p}) -- (\ref{omp}) to zero and solving the resulting polynomial equations for $x_1, x_2, x_3$ and $\Om_b$. The obtained solutions are the fixed points of the system denoted by variables with subscript ${}_f$, e.g., $\xc, \yc$ and $\zc$ are the fixed points for $x_1, x_2$ and $x_3$ respectively. It follows from eqs.~(\ref{addot}) and (\ref{omp}) that the fixed points which correspond to the acceleration of the universe can exist if $\yc \neq 0$ and $\Om_{b\, f} = 0$. When $\yc \neq 0$, eq.~(\ref{x2p}) gives the following relation for the fixed points:
\be
\yc = 1 + \sqrt{\frac{2}{3}} \lambda_4 \xc + \xc^2\,.
\label{x2x1}
\ee
Inserting this relation into eq.~(\ref{x3p}), supposing that $\xc \neq 0$ and using the fact that both $d x_3 / dN$ and $d x_1 / dN$ vanish at fixed point, we get
\be
0 = \sqrt{6} \left(-\lambda_1 + \lambda_2 + \left(\lambda_3 + 1\right) \lambda_4\right) \xc \zc\,.
\label{x3const}
\ee
Applying eqs.~(\ref{x2x1}) and (\ref{x3const}) to eq.~(\ref{x1p}), we can compute the fixed points for the case where both $\xc$ and $\yc$ do not vanish. This case correspond to the scaling solution which will be analyzed in detail in section \ref{scl-sol}.

We note that the relation in eq.~(\ref{x3const}) is derived by supposing $\xc \neq 0$. In the case $\xc = 0$, eq.~(\ref{x2x1}) gives $\yc = 1$, i.e., this fixed point is the potential dominated solution. We will consider this case in detail in the next section.

\subsection{Potential dominated solution}
\label{pot-sol}

The potential dominated solution corresponds to the fixed point $(\xc, \yc) = (0,1)$. Substituting this fixed point into eq.~(\ref{x1p}) and setting $\Om_b = 0$, we get $\lm_4 = 0$ at the fixed point. Setting $\lm_4 = 0$, eq.~(\ref{x1p}) yields
\be
\lim_{\xc \to 0} \(\frac 1{\xc} \left. \frac{d x_1}{d N}\right|_{x_1 = \xc, x_2 = \yc, \Om_b = 0} \) = - 3\,.
\ee
Substituting this relation into eq.~(\ref{x3p}), we obtain
\be
-6 \lambda_3 \zc = 0\,.
\ee
This implies that the fixed point that corresponds to the potential dominated solution occurs in two situations. The first is the situation where the disformal coefficient is much smaller than the conformal coefficient, i.e., $\zc = 0$ The second is the situation where  the disformal coefficient do not depend on $X$, i.e., $\lm_3 = 0$. This result can be easily understood by noting that the field $\phi$ is nearly constant in time within the potential dominated regime, so that the ratio of disformal coefficient to conformal coefficient nearly vanish if the disformal coefficient depends on kinetic term of the scalar field.
Performing the usual stability analysis, one can show that the fixed point for the potential dominated solution is stable for $\lm_3 \geq 0$.

\subsection{Scaling and field dominated solutions}
\label{scl-sol}

We now  consider the case where both $\xc$ and $\yc$ do not vanish.
In our consideration, $\xc^2 + \yc \leq 1$, i.e., $\Om_c \geq 0$ at the fixed points, so that these fixed points correspond to scaling and field dominated solutions.

According to eq.~(\ref{x3const}), the existence of the fixed points requires $\zc = 0$ or
\be
\lambda_2 = \lambda_1 - \left(\lambda_3 + 1\right) \lambda_4\,.
\label{condition}
\ee
Since $\zc = 0$ implies that disformal coefficient vanishes at the fixed point,
this fixed point is the conformal scaling solution. Hence, the case $\zc \neq 0$ corresponds to the disformal scaling solutions, in which the condition given in eq.~(\ref{condition}) is required for the existance of fixed points. We will consider these fixed point in the following sections.

\subsubsection{Conformal scaling solutions}

Substituting eq.~(\ref{x2x1}) into eq.~(\ref{x1p}) and then setting $\zc = 0$, we obtain the following fixed points:
\ba
\(\xc,\yc,\zc\) 
&=&
\(-\frac{\lambda_4}{\sqrt{6}}, 1 - \frac{\lambda_4^2}{6},0\)\,,
\label{fx-conf1}
\\
\(\xc,\yc,\zc\) 
&=&
\(\frac{\sqrt{6}}{\lambda_1 - 2 \lambda_4}, \frac{\lambda_1^2-2 \lambda_4 \lambda_1+6}{\left(\lambda_1-2 \lambda_4\right){}^2},0\)\,.
\label{fx-conf2}
\ea
The first fixed point is actually the scalar field dominated solution because $\xc^2 + \yc = 1$, while the second fixed point is the scaling solution. Since $\Om_b$ always vanishes at the fixed point in our consideration, we will perform stability analysis by linearizing only eqs.~(\ref{x1p}) - (\ref{x3p}). The inclusion of the evolution equation for the baryon density given in eq.~(\ref{omp}) will give rise to additional eigenvalue, $\mu = 3 \xc^2 -3 \yc$, which is negative for the fixed points corresponding to accelerating universe. To simplify our analysis, we will write the eigenvalues of the fixed points in terms of the density parameter $\ode$ and equation of state parameter $\wde$ of dark energy at fixed point. In terms of the dynamical variables defined in eq.~(\ref{def-dyna}), the quantities $\ode$ and $\wde$ at fixed point can be written as
\be
\odc = \xc^2 + \yc\,,
\qquad
\wdc = \frac{\xc^2 - \yc}{\xc^2 + \yc}\,,
\label{odwdc}
\ee
where $\odc$ and $\wdc$ are the value of $\ode$ and $\wde$ at fixed points respectively.
Inserting the fixed points from eqs.~(\ref{fx-conf1}) and (\ref{fx-conf2}) into the above equations, we respectively get
\ba
&& \lm_4 = \mp \sqrt{3(1 + \wdc)}\,,
\mbox{for the fixed points in eq.~(\ref{fx-conf1})}\,,
\label{lm4-conf1}\\
&& \left.
\begin{array}{c}
\lm_1 = \mp \frac{2 \wdc\sqrt{3\odc} }{\sqrt{1 + \wdc}}\,, \\
\lm_4 = \mp\frac{\sqrt{3} \left(1 + \wdc\odc\right)}{\sqrt{(1 + \wdc)\odc}}\,,
\end{array} \right\}
\mbox{for the fixed points in eq.~(\ref{fx-conf2})}\,. \nonumber \\
\label{lm4-conf2}
\ea
The eigenvalues $\mathbf{\mu} \equiv (\mu_1,\mu_2,\mu_3)$ for the fixed points in \\
eqs.~(\ref{fx-conf1}) and (\ref{fx-conf2}) are respectively given by
\ba
\mathbf{\mu} &=& \Big(- 3 (\lambda_3 + 1) \gac \mp \sqrt{3\gac} (\lambda_1 - \lambda_2), \nonumber\\
&& - \frac{3}{2} (1 - \wdc),
3 \wdc \pm \frac{1}{2} \sqrt{3\gac} \lambda_1 \Big)\,, 
\label{ei-fd}\\
\mathbf{\mu} &=& \Big(-6 + 3 (1 - \lambda_3) (1 + \wdc \odc) \pm \lambda_2 \sqrt{3 \gac \odc}, \nonumber\\
&& - \frac{3}{4} (1 - \wdc \odc + \frac{A}{\sqrt{\gac}}),- \frac{3}{4} (1 - \wdc \odc \nonumber\\
&& - \frac{A}{\sqrt{\gac}})\Big)\,,
\label{ei-sc}
\ea
where $\gac \equiv 1 + \wdc$ and
\ba
A \equiv && \big [\(1 - \wdc\) - 2 \(1 - \odc\) \(4 - 5 \wdc\) + \gac \wdc^2 \odc^2 \nonumber\\
&& - 2 \wdc^2\odc \big]^{1/2}\,.
\label{a-conf}
\ea
The stabilities of the fixed points in eqs.~(\ref{fx-conf1}) and (\ref{fx-conf2}) have been already discussed in literature \cite{conformc, stab:conform}, so that we will not consider in detail here. However, we would like to check whether the values for parameters $\lm_4, \lm_3, \lm_2, \lm_1$ can imply the evolution of the universe at late time using the dynamical analysis. Before considering more complicate fixed points in the next sections, let us start with the potential dominated solution discussed in section \ref{pot-sol} and conformal scaling solutions given in eqs.~(\ref{fx-conf1}) and (\ref{fx-conf2}). In the calculation for the conformal scaling solutions, we suppose that $\xc \neq 0$, so that $\wdc > -1$ and therefore
$\lm_4 = 0$ is not allowed in eqs.~(\ref{lm4-conf1}) and (\ref{lm4-conf2}). This implies that the universe will evolve towards the potential dominated solution, corresponding to the De Sitter expansion, at late time if $\lm_4 = 0$ and $\lm_3 \geq 0$.

In the case $\lm_4 < 0$, the universe evolves towards the stable fixed point $(\wde,\ode) = (\wdc,1)$ if $\lm_1 < - 6 \wdc / \sqrt{3\gac}$ and $\lm_3$ as well as $\lm_2$ are suitably chosen, e.g. $\lm_3 \sim \lm_2 \sim {\cal O}(1)$.
Similarly, for the case $\lm_4 > 0$, the universe will evolve towards the stable fixed point $(\wde,\ode) = (\wdc,1)$ if $\lm_1 > 6 \wdc / \sqrt{3\gac}$.
Here, $\wdc$ can be specified by $\lm_4$ through eq.~(\ref{lm4-conf1}).
However, if $\lm_4$ and $\lm_1$ satisfy eq.~(\ref{lm4-conf2}) and $\lm_3$ as well as $\lm_2$ are suitably chosen,
the universe will reach the stable fixed point $(\wde, \ode) = (\wdc, \odc)$ at late time, where $\wdc$ and $\odc$ are related to $\lm_1$ and $\lm_4$ through eq.~(\ref{lm4-conf2}).
We note that if we set $\lm_1, \lm_2, \lm_3$ and $\lm_4$ such that eq.~(\ref{condition}) is satisfied, the first eigenvalue in eqs. (\ref{ei-fd}) and (\ref{ei-sc}) will be zero. Consequently, one can show that these fixed points are saddle points, and therefore the universe will finally evolve towards the disformal scaling solutions discussed in the next section. Hence, in this section, we will consider only the cases where the relation in eq.~(\ref{condition}) is not satisfied.

It is interesting to check whether the fixed points in \\
eqs.~(\ref{fx-conf1}) and (\ref{fx-conf2}) can be stable for the same value of $\lm_1, \lm_2, \lm_3$ and $\lm_4$, i.e., the universe has two possible  stable fixed points for the same value of parameters. Let us suppose that $\wdc$ for the fixed point in eq.~(\ref{fx-conf1}) is $\wdc{}_1$, so that eq.~(\ref{lm4-conf1}) gives $\lm_4 = \mp \sqrt{3(1 + \wdc{}_1)}$. We then set $(\wdc, \odc) $\\
$= (\wdc{}_2, \odc{}_2)$ for the fixed point in eq.~(\ref{fx-conf2}), and use eq.~(\ref{lm4-conf2}) to show that 
\ba
\lm_4 = \mp \frac{\sqrt{3} (1 + \wdc{}_2 \odc{}_2)}{\sqrt{(1 + \wdc{}_2) \odc{}_2}} \nonumber\,,
\ea
for this fixed point. In the case where $\lm_4$ computed from both fixed points are the same, we can write $\odc{}_2$ in terms of $\wdc{}_1$ and $\wdc{}_2$ as
\begin{align}
\odc{}_2  =& \Big[1 + \wdc{}_1 - \wdc{}_2 + \wdc{}_1 \wdc{}_2 \nonumber\\
& \pm \Big( (1 + \wdc{}_1) (1 + \wdc{}_2) (\wdc{}_1 \wdc{}_2 + \wdc{}_1 \nonumber\\
& - 3 \wdc{}_2 + 1) \Big)^{1/2} \Big] / (2 \wdc{}_2^2)\,.
\label{odc2}
\end{align}
Applying a simple numerical calculation to the above equation, it can be checked that both values of $\odc{}_2$ are unphysical unless $-0.99 \leq \wdc{}_2 < \wdc{}_1 \leq 1$. 
Hence, if this condition on $\wdc{}_1$ and $\wdc{}_2$ is satisfied and $\lm_1$ as well as $\lm_4$ satisfy eq.~(\ref{lm4-conf2}), the universe will evolve towards the fixed point $(\wdc, \odc) = (\wdc{}_2, \odc{}_2)$ because this fixed point is stable.
The universe will not evolve to fixed point given in eq.~(\ref{fx-conf1}) because this fixed point becomes unstable in this situation.
To show that the fixed point in eq.~ (\ref{fx-conf1}) is unstable in this case,
we compute $\lm_1$ from eq.~(\ref{lm4-conf2}) by setting $(\wde,\ode) = (\wdc{}_2, \odc{}_2)$ and then insert the result into the third eigenvalue in eq.~(\ref{ei-fd}).
Therefore we get
\be
3 \wdc{}_1 - \frac{3 \wdc{}_2 \sqrt{\odc{}_2(1 + \wdc{}_1)}}{\sqrt{1+ \wdc{}_2}} > 0\,,
\label{thirdei}
\ee
for the case $-0.99 \leq \wdc{}_2 < \wdc{}_1 \leq 1$, $\odc{}_2$ satisfies eq.~(\ref{odc2}) and $0 < \odc{}_2 < 1$.
Nevertheless, if $\lm_1$ and $\lm_4$ do not satisfy eq.~(\ref{lm4-conf2}),
eq.~(\ref{thirdei}) is not valid
and hence the universe can evolve to the stable fixed point $(\wde, \ode) = (\wdc, 1)$ associated with eq.~(\ref{fx-conf1}).

 \subsubsection{Disformal scaling solutions}

In the case where the relation in eq.~(\ref{condition}) is satisfied,
the fixed points at which the disformal coefficient does not vanish, i.e., $\zc \neq 0$, can exist. Hence, these fixed points correspond to the disformal scaling solutions.
However,  since the LHS of eq.~(\ref{x3p}) vanishes due to the condition in eq.~(\ref{condition}) for this case,
eq.~(\ref{x3p}) does not provide relation among $\xc$, $\yc$ or $\zc$.
Therefore, we have only two relations among $\xc$, $\yc$ and $\zc$ obtained from eqs.~(\ref{x1p}) and (\ref{x2p}).
This suggests that we cannot solve the relations among $\xc$, $\yc$ and $\zc$ to write these fixed points completely in terms of the parameters of the model.
Hence, the values for one of the fixed point $\xc$, $\yc$ or $\zc$ can be chosen independently of the parameters of the model in our dynamical analysis.
To ensure that the chosen values of $\xc$, $\yc$ or $\zc$ are in agreement with the observational bounds,
we  write $\xc$ and $\lm_4$ in terms of $\wdc$ and $\odc$ by inserting eq.~(\ref{x2x1}) into eq.~(\ref{odwdc}) and solving the resulting equations as
\be
\xc = \pm \frac{\sqrt{(1 + \wdc)\odc}}{\sqrt{2}} \,, \hspace{0.1cm}
\lm_4 = \mp \frac{\sqrt{3} \left(1 + \wdc\odc\right)}{\sqrt{(1 + \wdc)\odc} }\,.
\label{x1lm4}
\ee
This implies that we can choose $\xc$ and $\lm_4$ by fixing $\wdc$ and $\odc$.
Substituting these relations into eq.~(\ref{x2x1}), we obtain
\be
\yc = \frac{1}{2} \left(1 - \wdc\right) \odc\,.
\label{x2c-dis}
\ee
Since the above equation is computed from eq.~(\ref{x2x1}),
this equation is a consequence of the vanishing of the LHS of eq.~(\ref{x2p}).

Substituting relations in eqs.~(\ref{condition}) and (\ref{x2x1}) into eq.~(\ref{x1p}), we obtain the polynomial equation which has degree 2 in $\zc$ and degree 6 in $\xc$. Since it is not easy to solve this equation for $\xc$ that lies inside the physical phase space, we instead solve this equation for $\zc$, and write $\xc$ and $\lm_4$ in the solutions in terms of $\wdc, \odc$ using eq.~(\ref{x1lm4}) as
\ba
\zc{}_1 &=&
\frac{1}{3 \lambda_3 \odc(1 +  \wdc) }\,,
\label{x3-sol1}\\
\zc{}_2 &=& \Big[ 6 \wdc \sqrt{\odc} \pm \lambda_1 \sqrt{3 (1 + \wdc)}\Big] / \Big[ 3 (1 + \wdc) \odc \nonumber\\
&& \times \big(6 (2 \lambda_3 + 1) \wdc \sqrt{\odc} \pm \lambda_1 \sqrt{3 (1 + \wdc)}\big) \Big]
\label{x3-sol2}
\ea
where $\zc{}_1$ and $\zc{}_2$ are the solutions for the polynomial equation of the fixed points. 
From the above calculations, we conclude that there are two classes of the fixed points for the disformal scaling solutions shown in table (\ref{classI-II}).
\begin{table}\begin{center}
\begin{tabular}{|l|l|l|l|}
\hline
class & $\xc$ & $\yc$ & $\zc$ \\
\hline
I
&
$\displaystyle{\pm \frac{\sqrt{(1 + \wdc)\odc}}{\sqrt{2}}}$
&
$\displaystyle{\frac{1}{2} \left(1 - \wdc\right) \odc}$
&
$\zc{}_1$
\\
\hline
II &
$\displaystyle{\pm \frac{\sqrt{(1 + \wdc)\odc}}{\sqrt{2}}}$
&
$\displaystyle{\frac{1}{2} \left(1 - \wdc\right) \odc}$
&
$\zc{}_2$
\\
\hline
\end{tabular}
\caption{\label{classI-II}
The fixed points for the disformal scaling solutions. For these fixed points, the parameters $\lm_1$ and $\lm_3$ are arbitrary, while $\lm_2$ and $\lm_4$ are replaced by eqs.~(\ref{condition}) and (\ref{x1lm4}) respectively.}
\end{center}\end{table}
It follows from eq.~(\ref{x3-sol1}) that $\zc{}_1 \to \infty$ when $\lm_3 = 0$, implying that the class I of fixed point does not exist for the case where the disformal coefficient does not depend on the kinetic terms of scalar field. It can be checked that for the case $\lm_3 = 0$, the fixed points belonging to the class II are the fixed points discussed in \cite{Sakstein:141}. According the result in \cite{Sakstein:141}, one of the eigenvalues for each fixed points in the class II is zero for the case where $\lm_3 = 0$. Hence, to simplify our analysis, we check whether the fixed points in the both classes have zero eigenvalue. We compute the metric $M_{ij} \equiv \pa E_i / \pa x_j$, where the indices  $i$ and $j$ run from 1 to 3 and $E_i$ is the RHS of eqs.~(\ref{x1p}) -- (\ref{x3p}) respectively. Evaluating this matrix at the fixed points, we get
\ba
\det M &&= \frac{\partial E_1}{\partial x_3} \big[
3 (\lambda_3 + 1) \big(\sqrt{6} \lambda_4 + 6 \xc \big) \xc \yc \nonumber\\
&& + \big(9 (\lambda_3 + 1) \xc^2 + 2 \sqrt{6} (\lambda_1 - \lambda_2) \xc \nonumber\\
&& -3 (\lambda_3 + 1) (\yc - 1)\big) \big(\sqrt{6} \lambda_4 \xc + 3 \xc^2 - 6 \yc \nonumber\\
&& + 3\big)\big] \zc
\ea
Inserting the fixed points in table~(\ref{classI-II}) into this matrix,
we get $\det M = 0$ for all fixed points which suggests that one of the eigenvalues for each fixed points vanishes. Therefore, to determine the stability of the fixed points, we have to go beyond linear analysis.
However, as discussed above for the case of disformal solution, we have only two relations among $\xc$, $\yc$ and $\zc$ which are not sufficient for writing the fixed points completely in terms of the parameters of the model.
Hence, is interesting to check whether the constraint equation given in eq.~(\ref{condition}) can imply the constraint equation for the dynamical variables $x_1$, $x_2$ and $x_3$.
Using the relation in eq.~(\ref{condition}) and the definitions of $x_1, x_2$ and $x_3$, we get
\be
\zc = \frac 1{3 C_0}\(\frac{M_v}{M}\)^{4 + 4\lm_3} \frac{\xc^{2 \lm_3}}{\yc^{1 + \lm_3}}
=  r_0 \frac{\xc^{2 \lm_3}}{\yc^{1 + \lm_3}}\,,
\label{x3reduce}
\ee
where $r_0$ is constant which controls magnitude of $x_3$ for given $x_1$ and $x_2$.
Hence, the third relation among $\xc$, $\yc$ and $\zc$ can be constructed from constraint given in eq.~(\ref{condition}) by introducing other constant parameter $r_0$.
The above constraint equation suggests that for a given $r_0$, $x_3$  can be computed from $x_1$ and $x_2$,
so that the dimension of the phase space is reduced \cite{Sakstein:141}.
Substituting this relation and the relation in eq.~(\ref{x2x1}) into eq.~(\ref{x1p}), we obtain polynomial equation degree $6 + 4 \lm_3$ of $\xc$. Similar to the previous analysis, instead of solve this equation for $\xc$, we solve this equation for $r_0$, so that we get two expressions for $r_0$:
\ba
r_0 &=& r_{01} = \frac{1}{6 \lambda_3} \(\frac{1 - \wdc}{1 + \wdc}\)^{\lm_3 + 1}\,,
\label{r01}\\
r_0 &=& r_{0\pm} = \frac{1}{2 \sqrt{3}} (\frac{1 - \wdc}{1 + \wdc})^{\lm_3 + 1}\nonumber\\
&& \times \frac{\lambda_1 \sqrt{1 + \wdc} \pm 2\wdc \sqrt{3\odc}} {\lambda_1 \sqrt{3(1+ \wdc)} \pm 6 \left(2 \lambda_3+1\right) \sqrt{\odc} \wdc}\,.
\label{r02}
\ea
Moreover, we also use the relations in eq.~(\ref{x1lm4}) to write the above expressions in terms of $\wdc$ and $\odc$.
The above relations suggest that For any given $\wdc$, $\odc$ and the parameters of the model,
the LHS of eq.~(\ref{x3p}) vanishes (fixed point exists) if $r_0$ satisfies the above expressions.
Here, $\xc$  and $\lm_4$ are presented in terms of $\wdc$ and $\odc$,
therefore the different values of $\wdc$ and $\odc$ corresponds to different values of $\xc$, $\lm_4$ as well as $\yc$ (according to eq.~(\ref{x2c-dis})).
In terms of $r_0$, the fixed points in table~(\ref{classI-II}) now can be presented in table (\ref{class-redu}).

\begin{table}\begin{center}
\begin{tabular}{|l|l|l|l|}
\hline
class I: & $\xc$ & $\yc$ &  $r_0$ \\
\hline
$\ip$
&
$\displaystyle{\frac{\sqrt{(1 + \wdc)\odc}}{\sqrt{2}}}$
&
$\displaystyle{\frac{1}{2} \left(1 - \wdc\right) \odc}$
&
$r_{01}$
\\
$\im$
&
$\displaystyle{- \frac{\sqrt{(1 + \wdc)\odc}}{\sqrt{2}}}$
&
$\displaystyle{\frac{1}{2} \left(1 - \wdc\right) \odc}$
&
$r_{01}$
\\
\hline
class II: & $\xc$ & $\yc$ &  $r_0$ \\
\hline
$\iip$
&
$\displaystyle{\frac{\sqrt{(1 + \wdc)\odc}}{\sqrt{2}}}$
&
$\displaystyle{\frac{1}{2} \left(1 - \wdc\)\odc}$
&
$r_{0+}$
\\
$\iim$
&
$\displaystyle{- \frac{\sqrt{(1 + \wdc)\odc}}{\sqrt{2}}}$
&
$\displaystyle{\frac{1}{2} \left(1 - \wdc\)\odc}$
&
$r_{0-}$
\\
\hline
\end{tabular}
\caption{\label{class-redu}
The fixed points for the disformal scaling solutions. For these fixed points, the parameters $\lm_1$ and $\lm_3$ are arbitrary, while $\lm_2$, $\lm_4$ and $r_0$ are replaced by eqs.~(\ref{condition}), (\ref{x1lm4}) and (\ref{r01}) -- (\ref{r02}) respectively.}
\end{center}
\end{table}

\begin{figure}
\includegraphics[width=8cm, height=4cm]{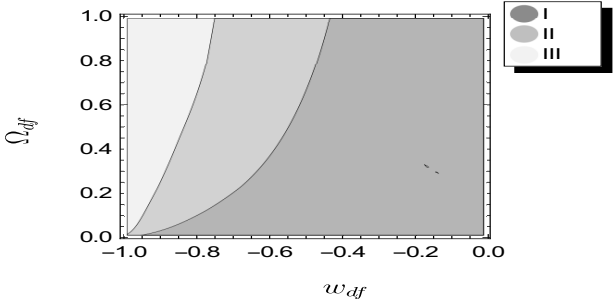}
\caption{\label{fig:1}
The regions I, II and III  represent the regions in which the fixed point $\iip$ is a saddle point for the cases where the values of $(\lm_3,\lm_1)$ are $(0,2)$, $(0,5)$ and $(0,10)$ respectively. The fixed point is stable outside these regions. Note that the region II totally overlaps with a part of region III, and the region I totally overlaps with a part of region II.}
\end{figure}

\begin{figure}
\includegraphics[width=8cm, height=4cm]{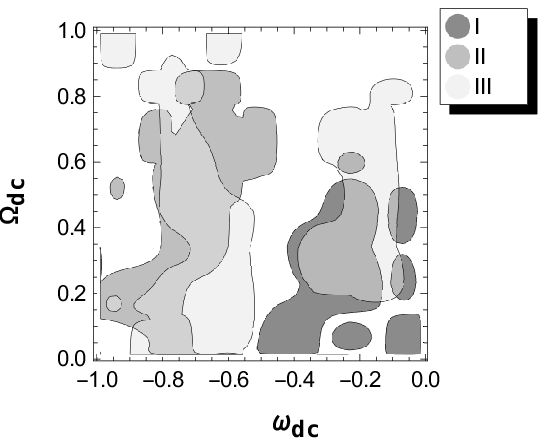}
\includegraphics[width=8cm, height=4cm]{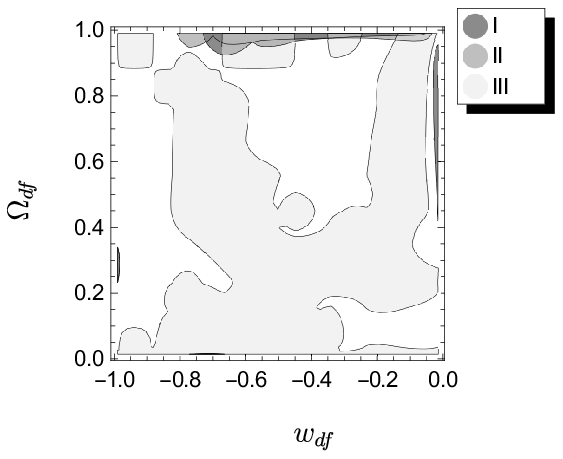}
\caption{\label{fig:2}
In the upper (lower) panel, regions I, II and III  represent the regions in which the fixed point $\iip$ ($\iim$) is a saddle point for the cases where the values of $(\lm_3,\lm_1)$ are $(1,1)$, $(1,5)$ and $(5,1)$ respectively. The fixed point is stable outside these regions.}
\end{figure}

\begin{figure}
\includegraphics[width=8cm, height=4cm]{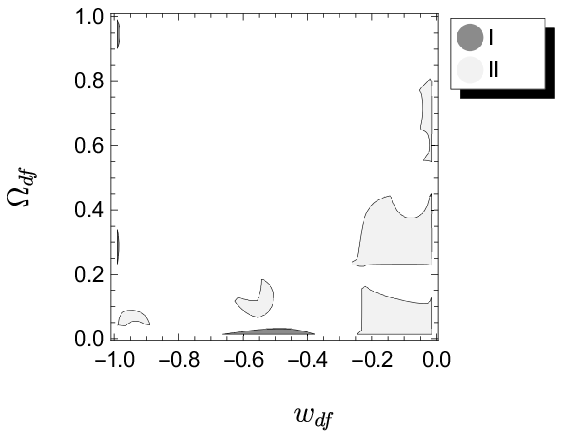}
\includegraphics[width=8cm, height=4cm]{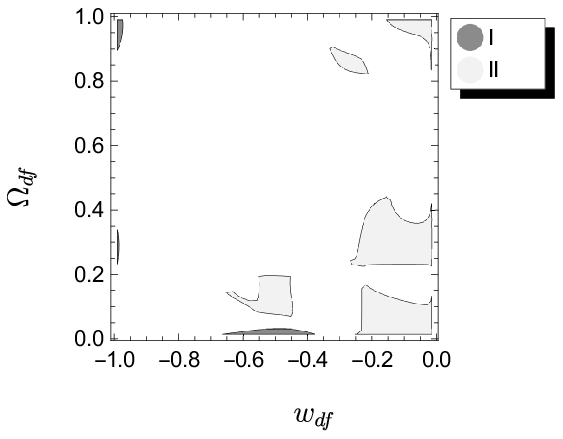}
\caption{\label{fig:3}
In the upper panel, regions I and II denote the stable  regions for the fixed point $\ip$ for the cases where the values of $(\lm_3,\lm_1)$ are $(1,100)$ and $(20,1)$ respectively. In the lower panel, regions I and II denote the stable  regions for the fixed point $\im$ for the cases where the values of $(\lm_3,\lm_1)$ are $(1,500)$ and $(20,1)$ respectively. The fixed point is saddle outside these regions.}
\end{figure}

\begin{figure}
\includegraphics[height=4cm, width=8.4cm]{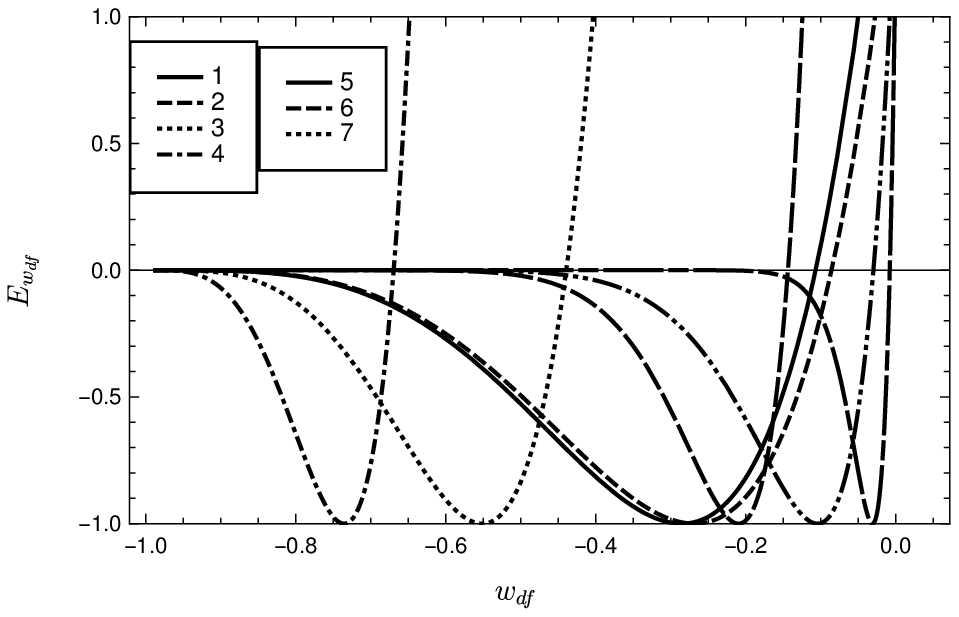}
\includegraphics[height=4cm, width=8.4cm]{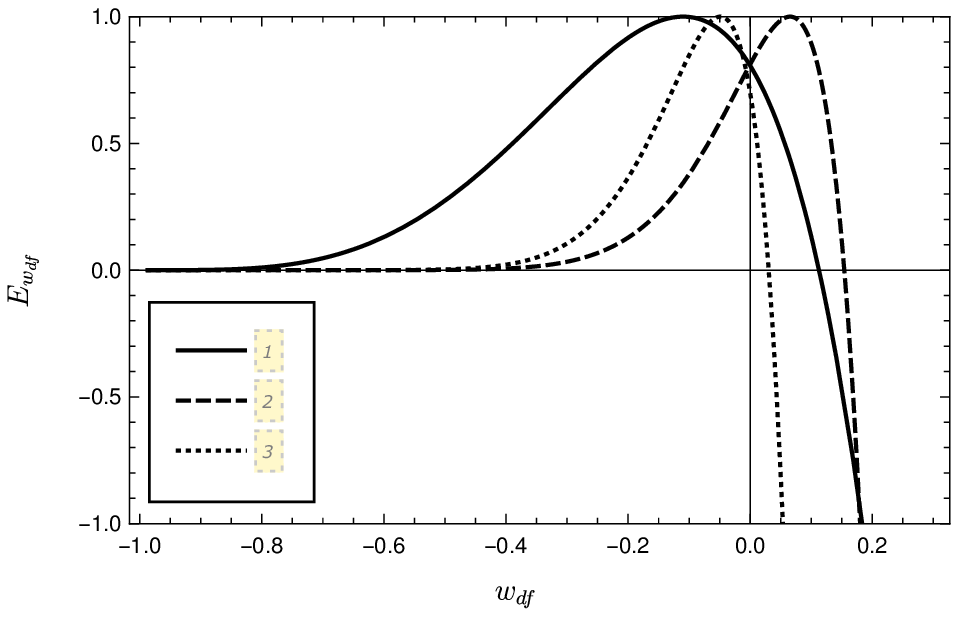}
\caption{\label{fig:4}
The plots of $E_{\wdc}$ as a function of $\wdc$ for the fixed points in the class $\iip$ (upper panel) and $\iim$ (lower panel). In the upper panel, lines 1 -- 7 represent the case where $(\lm_1, \lm_3) =$ (1,1), (1,1), (5,1), (10,1), (5,5), (1,5) and (1,20) respectively. For line 2, $r_0$ and $\lm_4$ are computed by setting $\wdc = -0.98$ and $\odc = 0.9$,
while these parameters are computed by setting $\wdc = -0.99$ and $\odc = 0.9$ for the other lines. In the lower panel, lines 1 -- 3 represents the case where $(\lm_1, \lm_3) =$ (1,1), (5,5) and (5,5) respectively. For this panel, $r_0$ and $\lm_4$ are computed by setting $\wdc = -0.99$ and $\odc = 0.9$.}
\end{figure}

It can be checked that the fixed points in the class II reduce to the fixed points in \cite{Sakstein:141} when $\lm_3 = 0$.
Since the analytic expressions for the above fixed points are  complicate, we perform the stability analysis using numerical calculation and show the region of the cosmological parameters at fixed point $(\wdc, \odc)$ in which the fixed point is stable in figures (\ref{fig:1}) -- (\ref{fig:4}). In our consideration, $\lm_4$, $\xc$ and $\yc$ can be computed from $\wdc$ and $\odc$, while $\lm_2$ and $r_0$ can be computed from $\wdc$, $\odc$, $\lm_1$ and $\lm_3$. Therefore, the stability of the fixed points for the disformal scaling cases can be explored by plotting the stability regions of the fixed points in the $\wdc$ -- $\odc$ plane for various values of $\lm_1$ and $\lm_3$.

Let us first consider the case $\lm_3 = 0$. It is clear that  this case is not allowed for the fixed points in the class I. The numerical investigation shows that the fixed point $\iim$ is stable for a wide range of $\lm_1$ when $-1 < \wdc < 0$ and $0 < \odc < 1$. However, it follows from figure (\ref{fig:1}) that the parameters region in which the fixed point $\iip$ is saddle increases area with increasing $\lm_1$. The fixed point $\iip$ can become a saddle point within the region $\wdc \in (-0.99, -0.97)$ and $\odc \in [0.7,1)$ until $\lm_1 \gg 1$  and $\lm_3 \gg 1$.

We now turn to the case $\lm_3 > 1$. It follows from figure (\ref{fig:2}) that for the fixed points in class $II$, the area of the saddle region in the $\wdc$ -- $\odc$ plane increases when  $\lm_3$ or $\lm_1$ increases. Similar to the case $\lm_3 = 0$, the fixed points in the class II can become a saddle points within the region $\wdc \in (-0.99, -0.97)$ and $\odc \in [0.7,1)$ unless $\lm_1 \gg 1$.

The fixed points in the class I are saddle points for the whole region of $-1 < \wdc < 0$ and $0 < \odc < 1$ when $\lm_1 = 1$ and $\lm_3 = 1$.
However, according to figure (\ref{fig:3}), these fixed points become stable within some regions in the $\wdc$ -- $\odc$ plane when $\lm_1$ or $\lm_3$ increases. For this class, the fixed points can be stable within the region $\wdc \in (-0.99, -0.97)$ and $\odc \in [0.7,1)$ if $\lm_1$ is sufficiently large, i.e., the slope of the conformal coefficient is large.

We now consider whether the same value of parameters $\lm_1, \lm_2, \lm_3$ and $\lm_4$ can lead to different stable fixed point at late time. Since the fixed points in the class I do not support $\lm_3 = 0$, we consider the case where $\lm_3 = 0$ for the fixed points in the class II. In the case where $\lm_3 = 0$ , eq.~(\ref{r02}) becomes
\be
r_{0\pm} = \frac 16 \frac{1 - \wdc}{1 + \wdc}\,.
\label{r0lm30}
\ee
This shows that for a given value of $r_0$, or equivalently $\zc$, the above equation can be satisfied by single value of $\wdc$. Inserting the value of $\wdc$ computed from eq.~(\ref{r0lm30}) into \\
eq.~(\ref{x1lm4}), we see that the values of $\odc$ that satisfy eq.~(\ref{x1lm4}) for a fixed value of $\lm_4$ are given by
\be
\sqrt{\odc} =
\left\{ 
\begin{array}{ll}
\frac{- \lm_4\sqrt{1 + \wdc} \pm \sqrt{(1 + \wdc)\lm_4^2 - 12 \wdc}}{2 \sqrt{3} \wdc}
& \mbox{for negative $\lm_4$} \\
- \frac{- \lm_4\sqrt{1 + \wdc} \pm \sqrt{(1 + \wdc)\lm_4^2 - 12 \wdc}}{2 \sqrt{3} \wdc}
& \mbox{for possitive $\lm_4$} \\
\frac{\sqrt{3}}{|\lm_4|}
& \mbox{for}\quad \wdc = 0\,. \\
\end{array}\right.
\label{sqrto-lm30}
\ee
This equation shows that $\odc$ takes a single value for a given $\lm_4$ if $\wdc = 0$. However, if $\wdc \neq 0$, there are two possible values of $\odc$ for a given $\lm_4$. When $-1 < \wdc < 0$ and $\lm_4$ is a real number, it follows from the above equation that, for a given value of $\lm_4$ and $\wdc$, one possible value of $\sqrt{\odc}$ is positive while the another is negative. For $0 < \wdc < 1$, the numerical evaluation of eq.~(\ref{sqrto-lm30}) shows that if the values of $\lm_4$ and $\wdc$ are chosen such as the first value of $\odc$ lies within the range $(0,1)$,
the second value of $\odc$ will be larger than unity. According to these analysis, only single fixed point lies inside the physical phase space for a given value of $\lm_1, \lm_3,\lm_4$ and $r_0$ when $\lm_3 = 0$.

In the cases where $\lm_3 \neq 0$ and $-1 < \wdc < 1$, eq.~(\ref{r01}) suggests that $\wdc$ takes a single value for a given real value of $r_0$ and $\lm_3$. Using the similar analysis as in the case where $\lm_3 = 0$,
we conclude that the fixed points $(\wdc, \odc)$ belonging to the class I take single physically relevant value for a given value of $\lm_1, \lm_3, \lm_4$ and $r_0$.

The situation changes when we consider the fixed points in the class II.
It can be checked that for a given value of $r_0$ and the other parameters, eq.~(\ref{r02}) is satisfied by several values of $\wdc$ and $\odc$. The relation among $\wdc$ and the parameters of the model can be computed by combining eq.~(\ref{r02}) with eq.~(\ref{x1lm4}), so that we get
\ba
E_{\wdc} &&=
\sqrt{3} \Big[(1 - \wdc)^{2 + 2\lm_3} \big(12 \wdc + \lm_1^2 (1 + \wdc)\nonumber\\
&& - 2 \lm_1 \lm_4 (1 + \wdc)\big) + 36 r_0^2 \(1 + \wdc\)^{2 + 2 \lm_3}  \nonumber\\ 
&& \times \Big(12 (1 + 2 \lm_3)^2 \wdc + \lm_1^2 (1 + \wdc) \nonumber\\
&& - 2 \lm_1 \lm_4 (1 + 2\lm_3) (1 + \wdc)\Big) - 12 r_0 \(1 - \wdc^2\)^{1 + \lm_3} \nonumber\\
&& \times \Big(\lm_1^2 (1 + \wdc) - 2\lm_1 \lm_4 (1 + \lm_3) (1 + \wdc)\nonumber\\
&& + 12 \wdc (1 + 2 \lm_3 )\Big)\Big] = 0 \,.
\label{ewdc}
\ea
to compute values of $\wdc$ that satisfy the above equation for a given value of $\lm_1,\lm_3,\lm_4$ and $r_0$, we perform the following analysis.
We first use eqs.~(\ref{r02}) and (\ref{x1lm4}) to compute $\lm_4$ and $r_0$ for selected value of $\lm_1, \lm_3, \wdc$ and $\odc$. Since the disformal fixed point exists when the energy density of baryon can be neglected compared with the energy density of dark matter, this fixed point should be reached in the future. Hence, the value of $\wdc$ and $\odc$  are chosen such that the present value of $w_d$ and $\Om_d$ can be in agreement with the observational bounds. We then substitute the computed value of $\lm_4$ and $r_0$ as well as the selected value of $\lm_1$ and $\lm_3$ into eq.~(\ref{ewdc}). It is clear that the value of $\wdc$ that is used to compute $\lm_4$ and $r_0$ is the solution of the resulting equation. In the following consideration, we will see that there is the other solution for this equation corresponding to the stable physically relevant fixed point. This implies that there are two stable physically relevant fixed point for the same value of the parameters of the model.

For illustration, we plot eq.~(\ref{ewdc}) in figure (\ref{fig:4}) for the case where $r_0$ and $\lm_4$ are computed from $\wdc = \wdc^* = -0.99$, $-0.98$ and $\odc = \odc^* = 0.9$. From the plots, we see that in addition to the obvious solution $\wdc = \wdc^*$, eq.~(\ref{ewdc}) also has a solution in a range $\wdc = \wdc^s\in (-0.98, 0.15)$. In the cases where $\lm_1 \sim {\cal O}(1)$ and $\lm_3 \sim {\cal O}(1)$, the plots show that this solution lies within the ranges $\wdc^s \in (0,0.15)$ and $\wdc^s \in (-0.98,0)$ for the fixed points in the classes $\iim$ and $\iip$ respectively. The value of $\odc$ associated with $\wdc^s$ can be obtained by inserting $\wdc^s$ into eq.~(\ref{x1lm4}). for the ranges of parameters considered in figure (\ref{fig:4}), the values of $\odc$ lies within the range $(0,1)$. Applying the above stability analysis to the two fixed points associated with the zero points of $E_{\wdc}$ in figure (\ref{fig:4}), we find that both of the fixed points are stable fixed points. These results indicate that in the case where $\lm_3 > 0$, the fixed point $\iip$ (an also $\iim$) can take two different physically relevant values for the same value of the parameters of the model, and the fixed point is stable at these values.

It follows from figure (\ref{fig:4}) that for the fixed point $\iip$,
the solution $\wdc^s$ for $E_{\wdc} = 0$ shifts to the lower value when $\lm_1$ increases, and shifts to the larger value when $\lm_3$ increases.
According to our numerical investigation, we also find that $\wdc^s$ gets closer to 0 as $\lm_3$ gets larger, but $\wdc^s$ will not exist when $\lm_1 \gtrsim 30$. This means that the fixed point $\iip$ can take only one physically relevant value for a given value of the parameters when $\lm_1 \gtrsim 30$. From figure (\ref{fig:4}), we see that for the fixed point $\iim$, the solution $\wdc^s$ shifts towards 0 when $\lm_3$ increases. From the detail of the numerical analysis for fixed point $\iim$, we find that $\odc$ associated with $\wdc^s$ becomes larger than unity when $\lm_1 \gtrsim 1$. Nevertheless, the value of $\odc$ can be reduce by enhancing the value of $\lm_3$, e.g., $\odc < 1$ for $\lm_1 = \lm_3 = 5$. For both $\iip$ and $\iim$ fixed points, the solution $\wdc^s$ does not exist if $\odc^* \gtrsim 0.9$. Based on the above analysis, we conclude that  the fixed point associated with $\wdc^s$ will not exist if $\lm_1$ is sufficiently larger than unity or the values of $r_0$ and $\lm_4$ correspond to $\odc^* \gtrsim 0.9$.

\begin{figure}
\includegraphics[height=4cm, width=8.4cm]{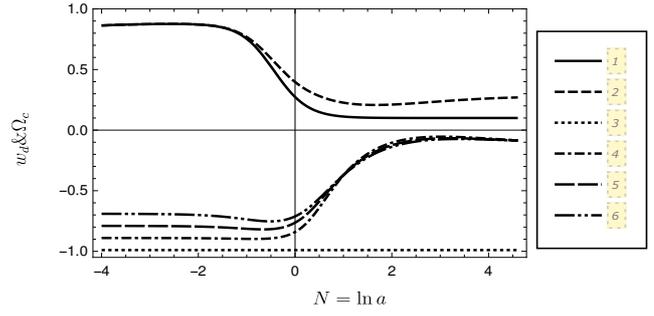}
\caption{\label{fig:5}
The evolution of $\Om_\dm$ and $\wde$. The lines 3 -- 6 represent the case where the initial conditions are setting such that the initial value of $\wde = \wdc^*, -0.89, - 0.79$ and $-0.69$ respectively. The first and second lines represent the evolution of $\Om_\dm$ for the case where the initial conditions for $\wde$ are equal to that for the line 3 and 4 respectively. Note that the evolution of $\Om_\dm$ is nearly the same for the case where the initial values of $\wde$ are $-0.89, -0.79$ and $-0.69$.}
\end{figure}

We now study the situation in which the different fixed points with the same value of the parameters of the model can be reached. In order to perform, we solve eqs.~(\ref{x1p}) -- (\ref{omp}) numerically by setting the present value of $\Om_\dm, \Om_b, \Om_r, \Om_d$ and $w_d$ to be $0.27, 0.03, 10^{-4}, 0.7 - 10^{-4}$ and $-0.99$ respectively. For illustration, we plot in figure (\ref{fig:5}) the evolution of $\Om_\dm$ and $\wde$ for the case where $\lm_1 = \lm_3 = 1$ and $r_0, \lm_4$ are computed from $(\wdc,\odc) = (\wdc^*,\odc^*) = (-0.99,0.9)$. 
From the figure, we see that if the initial conditions are chosen such that the initial value of $\wde$ is significantly larger than $\wdc^*$, the universe will evolve towards the fixed point associated to the solution $\wdc^s$ in figure (\ref{fig:4}). Unfortunately, the present value of $\wde$ may lie outside the observational bounds if the universe evolves towards this fixed point. Hence, the existence of the solution $\wdc^s$ seems to be a problem, which can be avoided by setting $\lm_1$ to be sufficiently larger than unity or setting the value of $r_0$ and $\lm_4$ to be matched with $\odc^* \gtrsim 0.9$. We stress that this conclusions are based on the situation where $\lm_3 > 0$ and $\lm_1 > 0$.

\section{Conclusions}
\label{sec:5}

In this work, we study dynamics of the universe at late time for the model in which dark energy directly interacts with dark matter through disformal coupling. When the disformal coefficient depends on the kinetic terms of scalar field, there exist two classes of fixed point which  can describe the acceleration of the universe in addition to that found in literature.
The fixed points in the first class are saddle points unless $\lm_1$ is sufficiently larger than unity, and exist only for the case where the disformal coefficient depends on the kinetic terms of scalar field. The fixed points in the second class can be stable within the parameters ranges that correspond to the accelerating universe.

In the case where the disformal coefficient depends only on the scalar field, the fixed points in the second class become the fixed points that found in the literature. Interestingly, in the case where the disformal coefficient also depends on the kinetic terms of scalar field, the stable fixed points in the second class can take different physically relevant values for the same value of the parameters of the model. For the case where $\lm_1 \sim \lm_3 \sim 1$ and the value of $r_0$ and $\lm_4$ are set such that the fixed point can occur at $0.9 \gtrsim \odc \gtrsim 0.7$ and $\wdc \sim -0.99$, the universe will evolve towards the fixed point $(\wde, \ode) = (\wdc,\odc)$ if the initial value of $\wde$ is close to $\wdc$. Nevertheless, if the initial value of $\wde$ is sufficiently larger than $\wdc$, the universe will evolve towards another value of the fixed point at which the present value of $\wde$ may not be in agreement with the observational bounds. The existence of two different values of the fixed point for the same value of the parameters can be avoided if $\lm_1$ is sufficiently larger than unity, or the values of $r_0$ and $\lm_4$ are set from the fixed point $\wdc \sim -0.99$ and $\odc \gtrsim 0.9$.

\begin{acknowledgements}
KK is supported by Thailand Research Fund (TRF) through grant RSA5780053.
\end{acknowledgements}

\end{document}